\documentclass[letter]{IEEEtran}
\usepackage[utf8]{inputenc} 
\usepackage[T1]{fontenc}
\usepackage{url}
\usepackage{ifthen}
\usepackage{cite}
\usepackage[cmex10]{amsmath} 
\usepackage{mathpple}
\usepackage{times}
\usepackage{amsthm,xpatch}
\usepackage{amsfonts}
\usepackage{amssymb}
\usepackage{breqn}
\usepackage{mathtools}
\usepackage{bbm}
\usepackage{comment}
\usepackage{multirow}
\usepackage{flushend}
\usepackage{float}

\newtheorem{theorem}{Theorem}

\newtheorem{remark}{Remark}

\newtheorem*{example*}{Example}

\renewcommand{\baselinestretch}{0.98}

\IEEEoverridecommandlockouts

\begin{document}

\pagestyle{plain}

\title{\fontsize{21}{28}\selectfont  A Low-Complexity Scheme for\\ Multi-Message Private Information Retrieval}

\author{Ningze Wang, Anoosheh Heidarzadeh, and Alex Sprintson\thanks{Ningze Wang and Alex Sprintson are with the Department of Electrical and Computer Engineering, Texas A\&M University, College Station, TX 77843  USA (E-mail: \{ningzewang, spalex\}@tamu.edu).} 
\thanks{Anoosheh Heidarzadeh is with the Department of Electrical and Computer Engineering, Santa Clara University, Santa Clara, CA 95053 USA (E-mail: aheidarzadeh@scu.edu).}
}

\maketitle 

\thispagestyle{plain}

\begin{abstract}
Private Information Retrieval (PIR) is a fundamental problem in the broader fields of security and privacy. In recent years, the problem has garnered significant attention from the research community, leading to achievability schemes and converse results for many important PIR settings. 

This paper focuses on the Multi-message Private Information Retrieval (MPIR) setting, where a user aims to retrieve \(D\) messages from a database of \(K\) messages, with identical copies of the database available on \(N\) remote servers. The user's goal is to maximize the download rate while keeping the identities of the retrieved messages private. Existing approaches to the MPIR problem primarily focus on either scalar-linear solutions or vector-linear solutions, the latter requiring a high degree of subpacketization. Furthermore, prior scalar-linear solutions are restricted to the special case of \(N =  D+1\). This limitation hinders the practical adoption of these schemes, as real-world applications demand simple, easily implementable solutions that support a broad range of scenarios.

In this work, we present a solution for the MPIR problem, which applies to a broader range of system parameters and requires a limited degree of subpacketization. In particular, the proposed scheme applies to all values of \(N=DL+1\) for any integer \(L\geq 1\), and requires a degree of subpacketization \(L\). Our scheme achieves capacity when \(D\) divides \(K\), and in all other cases, its performance matches or comes within a small additive margin of the best-known scheme that requires a high degree of subpacketization.
\end{abstract}

\section{Introduction}
\emph{Private Information Retrieval (PIR)} \cite{CGKS1995} is a fundamental problem in the fields of security and privacy, and in recent years, it has garnered significant attention from the research community.
This has led to the development of numerous achievability schemes and converse results for various key PIR scenarios. 
Early research primarily focused on \emph{multi-server single-message settings} (see e.g., \cite{SJ2017,TER2017,TGKHHER2017,TSC2019,ZTSP2021ISIT,LJJ2021}). 
Several subsequent works studied the design and analysis of efficient schemes for the \emph{single-server single-message} PIR settings in the presence of side information~(see e.g., \cite{KGHERS2020,HKS2019Journal,KHSO2021,LJ2022}). 

More recently, the focus has shifted to the \emph{Multi-message PIR (MPIR)} problem, due to its relevance to private computation and private machine learning. 
An instance of the MPIR problem includes a dataset of $K$ messages with a copy of this dataset stored at $N\geq 1$ remote servers that are not allowed to collude. 
A user seeks to retrieve $D > 1$ messages from the dataset while ensuring that the identities of the desired messages remain hidden from each server. 
The MPIR problem is motivated by several practical applications. 
For example, in a private machine-learning scenario, a user may need to retrieve a set of training points from a publicly available dataset while ensuring that the servers cannot determine which specific training points are requested. 
Similarly, the user might need to retrieve the current market value of a group of stocks before making an investment decision without disclosing the identities of these stocks to the servers. 

Several settings of the single-server MPIR problem were studied in~\cite{HKGRS2018,HKRS2019,KKHS32019,HS2022LinCap}. 
In~\cite{BU2018}, Banawan and Ulukus focused on the multi-server MPIR setting and proposed schemes that either achieve capacity (the maximal achievable download rate) in certain regimes or perform near capacity in others. 
However, the schemes in~\cite{BU2018} require dividing each message into a number of subpackets growing quadratically with $N$ when $D>\frac{K}{2}$, or exponentially with $K$ when $D\leq\frac{K}{2}$. 
This high degree of subpacketization introduces significant practical challenges, such as the need to index and account for every subpacket, resulting in a high volume of meta-data. Moreover, the large number of required subpackets imposes a requirement on the minimum message size and also leads to the need for high-complexity computations involving high-dimensional matrices during the decoding process.

In our previous work~\cite{WHS2022}, we focused on scalar-linear multi-server MPIR schemes, which do not utilize subpacketization and in which every query to the server is a linear combination of the original messages. 
These schemes offer significant advantages in practical settings due to their low implementation complexity. 
However, a key limitation of~\cite{WHS2022} is that it only addressed the case of ${N=D+1}$, leaving the design of a multi-server MPIR scheme---one that either does not require subpacketization or only requires a low subpacketization degree---for a broader range of system parameters an open problem.

In this work, we present a solution to the MPIR problem that applies to a broader range of system parameters and requires a degree of subpacketization that only grows linearly with $N$ and remains constant with respect to $K$. 
As a result, our scheme has low computational complexity, requires a small amount of meta-data, and is easy to implement in practical settings. 
In particular, we propose a vector-linear multi-server MPIR scheme that applies to all values of ${N=DL+1}$ for any integer $L\geq 1$, and requires a subpacketization degree of $L$. 
Our scheme achieves capacity when $D$ divides $K$. 
In all other cases, our scheme's download rate is either on par with the best previously-known achievable rates in~\cite{BU2018}, or differs by a small additive margin, being slightly higher in some cases and slightly lower in others.

\section{Problem Setup}\label{sec:SN}
We denote random variables and their realizations by bold-face and regular symbols, respectively. 
For any random variables $\mathbf{X}$ and $\mathbf{Y}$, we denote the Shannon entropy of $\mathbf{X}$ by $H(\mathbf{X})$ and the conditional entropy of $\mathbf{X}$ given $\mathbf{Y}$ by $H(\mathbf{X}|\mathbf{Y})$, both measured in bits. 
For any integer $i\geq 1$, we denote $\{1,\dots,i\}$ by $[i]$. 
For any integers $0\leq k\leq n$, we denote the binomial coefficient $\binom{n}{k}$ by $C_{n,k}$. 

Consider $N$ non-colluding servers, each storing an identical copy of $K$ messages $\mathrm{X}_1, \dots, \mathrm{X}_K$. 
Each message $\mathrm{X}_k \in \mathbbmss{F}_q^m$ for $k \in [K]$ comprises $m$ symbols, each an element of $\mathbbmss{F}_q$. 
Here, $\mathbbmss{F}_q$ is a finite field of order $q$, and $\mathbbmss{F}_{q}^{m}$ is the vector space of dimension $m$ over $\mathbbmss{F}_q$.

To simplify the notation, for every subset ${\mathrm{S}\subset [K]}$, we denote the collection of messages $\{\mathrm{X}_k: k\in \mathrm{S}\}$ by $\mathrm{X}_{\mathrm{S}}$. 
   
Consider a user who wishes to retrieve the set of $D$ messages $\mathrm{X}_{\mathrm{W}}$ for a given $\mathrm{W}\in \mathbbmss{W}$, where $\mathbbmss{W}$ is the set of all $D$-subsets of $[K]$.  
We refer to $\mathrm{X}_{\mathrm{W}}$ as the \emph{demand messages}, and $\mathrm{X}_{[K]\setminus \mathrm{W}}$ as the \emph{interference messages}. 

We assume that: 
(i) $\mathbf{X}_1,\dots,\mathbf{X}_K$ are independent and uniformly distributed over $\mathbbmss{F}_{q}^{m}$; 
(ii) $\mathbf{X}_{1},\dots,\mathbf{X}_K$ and $\mathbf{W}$ are independent; 
(iii) $\mathbf{W}$ is uniformly distributed over $\mathbbmss{W}$; 
and (iv) the distribution of $\mathbf{W}$ is known to all servers, whereas the realization $\mathrm{W}$ is not known to any server.

Given the demand's index set $\mathrm{W}$, the user generates a query $\mathrm{Q}_n^{[\mathrm{W}]}$ for each ${n\in [N]}$, and sends it to server $n$. 
Each query $\mathrm{Q}_n^{[\mathrm{W}]}$ is a deterministic or stochastic function of $\mathrm{W}$, and independent of $\mathrm{X}_1,\dots,\mathrm{X}_K$. 
For each $n\in [N]$, the query $\mathrm{Q}_n^{[\mathrm{W}]}$ must not reveal any information about the demand's index set $\mathrm{W}$ to server $n$. 
That is,
for every ${{\mathrm{W}^{*}}\in \mathbbm{W}}$, it must hold that
\begin{equation*}
\mathbb{P}(\mathbf{W}={\mathrm{W}^{*}}|\mathbf{Q}_n^{[\mathbf{W}]}=\mathrm{Q}_n^{[\mathrm{W}]})=\mathbb{P}(\mathbf{W}={\mathrm{W}^{*}}) \quad \forall n\in [N]. 
\end{equation*}
We refer to this condition as the \emph{privacy condition}. 

Upon receiving the query $\mathrm{Q}_n^{[\mathrm{W}]}$, server $n$ generates an answer $\mathrm{A}_n^{[\mathrm{W}]}$, and sends it back to the user. 
The answer $\mathrm{A}_n^{[\mathrm{W}]}$ is a deterministic function of $\mathrm{Q}_n^{[\mathrm{W}]}$ and $\mathrm{X}_1,\dots,\mathrm{X}_K$.
That is, 
\[H(\mathbf{A}_n^{[\mathrm{W}]}|\mathbf{Q}_n^{[\mathrm{W}]},\mathbf{X}_1,\dots,\mathbf{X}_K)=0.\] 

The user must be able to recover their demand messages $\mathrm{X}_{\mathrm{W}}$ given the collection of answers
${\mathrm{A}^{[\mathrm{W}]}:=\{\mathrm{A}_n^{[\mathrm{W}]}: n\in [N]\}}$, the collection of queries ${\mathrm{Q}^{[\mathrm{W}]}:=\{\mathrm{Q}_n^{[\mathrm{W}]}: n\in [N]\}}$, and the demand's index set $\mathrm{W}$. 
That is,
\[H(\mathbf{X}_{\mathrm{W}}| \mathbf{A}^{[\mathrm{W}]},\mathbf{Q}^{[\mathrm{W}]})=0.\] 
We refer to this condition as the \emph{recoverability condition}.

The problem is to design a protocol for generating a collection of queries $\mathrm{Q}^{[\mathrm{W}]}$ and the corresponding collection of answers $\mathrm{A}^{[\mathrm{W}]}$ for any given demand's index set $\mathrm{W}$ 
such that both the privacy and recoverability conditions are satisfied.
This problem, originally introduced in~\cite{BU2018} and subsequently studied in~\cite{WHS2022}, is referred to as \emph{Multi-message Private Information Retrieval (MPIR)}. 

In this work, we focus on (vector-) linear MPIR protocols with a subpacketization degree of $L$ (for $L\geq 1$), 
where each server's answer to the user's query consists only of linear combinations of message subpackets, with each message divided into $L$ subpackets, each comprising $m/L$ message symbols.   
That is, for each $n\in [N]$, the answer $\mathrm{A}^{[\mathrm{W}]}_n$ consists of one or more linear combinations of the message subpackets with combination coefficients from $\mathbbmss{F}_q$. 

We define the \emph{rate} of a protocol as the ratio of the number of bits required by the user, 
i.e., ${H(\mathbf{X}_{\mathbf{W}})}$, to the expected total number of bits downloaded from all servers, 
i.e., ${\sum_{n=1}^{N} H(\mathbf{A}^{[\mathbf{W}]}_n|\mathbf{Q}^{[\mathbf{W}]}_n)}$.
For any given $L$, we define the \emph{capacity} of linear MPIR as the supremum of rates over all linear MPIR protocols with a subpacketization degree of $L$.   

Our goal in this work is to design linear MPIR protocols that achieve capacity or rates close to capacity, with a low subpacketization degree $L$. 
In particular, we are interested in protocols that do not require $L$ to grow faster than linearly with $N$ or vary with $K$. 

\section{Main Results}\label{sec:main}
In this section, we present our main results, which apply to all values of $K$, $D$, and ${N=DL+1}$ for any $L\geq 1$.  

Let $\mathrm{M}$ be a $D\times D$ matrix defined as
\begin{equation}\label{eq:M}
\mathrm{M} := 
\begin{bmatrix} 
\frac{1}{\beta_1} & \frac{1}{\beta_1} & \cdots & \frac{1}{\beta_1} & \frac{1}{\beta_1} & \frac{1}{\beta_1}\\
\frac{\beta_1}{\beta_2} & 0 & \cdots & 0 & 0 & 0\\
0 & \frac{\beta_2}{\beta_3} & \cdots & 0 & 0 & 0\\
\vdots & \vdots & \ddots & \vdots & \vdots & \vdots\\
0 & 0 & \cdots & \frac{\beta_{D-2}}{\beta_{D-1}} & 0 & 0\\
0 & 0 & \cdots & 0 & \frac{\beta_{D-1}}{\beta_{D}} & 0\\
\end{bmatrix}, 
\end{equation} 
where $\beta_j:= DL/C_{D,j}$ for all $1\leq j\leq D$. 
Also, let ${\mathrm{f}=[f_1,\dots,f_D]^{\mathsf{T}}}$ and ${\mathrm{g}=[g_1,\dots,g_D]^{\mathsf{T}}}$ be column-vectors of length $D$ defined as
\begin{equation}\label{eq:f}
\mathrm{f}^{\mathsf{T}}:=\mathbf{1}^{\mathsf{T}}\mathrm{M}^{K-D},    
\end{equation}
and
\begin{equation}\label{eq:g}
\mathrm{g}^{\mathsf{T}}:=\mathbf{1}^{\mathsf{T}}(\mathrm{I}+\mathrm{M})^{K-D},    
\end{equation}
where $\mathbf{1}$ denotes the all-one column-vector of length $D$, and $\mathrm{I}$ denotes the $D\times D$ identity matrix.
 
\begin{theorem}\label{thm:1}
For linear MPIR with a subpacketization degree of $L$, a total of $K$ messages, $D$ demand messages, and ${N=DL+1}$ servers, the capacity is lower bounded by 
\begin{equation}\label{eq:R}
R=\frac{DL}{N-\max_{j\in [D]} f_j/g_j}, 
\end{equation} 
and upper bounded by 
\begin{equation}\label{eq:C}
\left(\frac{1-1/N^{\lfloor {K}/{D}\rfloor}}{1-1/N}+\left(\frac{K}{D}-\left\lfloor\frac{K}{D}\right\rfloor\right)\frac{1}{N^{\lfloor{K}/{D}\rfloor}}\right)^{-1}.
\end{equation} 
\end{theorem}

The upper bound in~\eqref{eq:C} is presented without proof and directly follows from the converse results in~\cite{BU2018}, which considers the supremum of rates over all MPIR protocols. 
To establish the lower bound, we propose a new linear MPIR scheme. 
In this scheme, each message is divided into $L$ ($=(N-1)/D$) equal-sized subpackets, and one linear combination of the message subpackets is retrieved from each server. 
A randomized algorithm is carefully designed to select the messages and subpackets for combination, ensuring that both privacy and recoverability are satisfied. 
This scheme generalizes the scheme we previously proposed in~\cite{WHS2022} which was limited to the special case of $L=1$.   

\begin{theorem}\label{thm:2}
For linear MPIR with a subpacketization degree of $L$, a total of $K$ messages, $D$ demand messages such that $D\mid K$, and ${N=DL+1}$ servers, the capacity is given by 
\vspace{-0.1cm}
\begin{equation}\label{eq:CC}
C=\frac{1-1/N}{1-1/N^{{K}/{D}}}.    
\end{equation}
\end{theorem}

The proof follows from the fact that when $D\mid K$, the lower bound in~\eqref{eq:R} and the upper bound in~\eqref{eq:C} are both equal to $C$ defined in~\eqref{eq:CC}. 

\begin{remark}\label{rem:1}\normalfont
Prior work~\cite{BU2018} presented a capacity-achieving MPIR scheme for $D\mid K$ and any $N$, which required dividing each message into a number of subpackets that grows exponentially with $K$. 
Interestingly, our results show that when $D\mid K$, capacity can also be achieved for any $N=LD+1$ with $L\geq 1$, with a subpacketization degree of $L$, which remains constant with respect to $K$. 
\end{remark}

\begin{remark}\label{rem:2}\normalfont
Our numerical results show that when ${D\nmid K}$ and ${D<\frac{K}{2}}$, the rate of our scheme is at least as high as that of the scheme in~\cite{BU2018}, and in some cases, it is higher, with the largest difference of approximately $0.000861$ for $K=5$, $D=2$, and $N=5$. 
In contrast, when ${D\nmid K}$ and ${D>\frac{K}{2}}$, our scheme achieves a lower rate than the scheme in~\cite{BU2018}, with the largest difference of approximately $0.044444$ for $K=5$, $D=4$, and $N=5$.
\end{remark}

\section{Proof of Theorem~\ref{thm:1}}
In this section, we propose a linear MPIR scheme that is applicable to all values of $K$, $D$, and $N=DL+1$ for any ${L\geq 1}$, and achieves the rate $R$ defined in~\eqref{eq:R}, thereby proving the lower bound in Theorem~\ref{thm:1}.

\subsection{Proposed Scheme}\label{subsec:Scheme}
The proposed scheme operates on message subpackets, where each message $\mathrm{X}_k$ is divided into ${L}$ subpackets, each composed of $m/{L}$ symbols from $\mathbbmss{F}_q$.
For each $k\in [K]$, let $\mathrm{X}_{k,1},\dots,\mathrm{X}_{k,L}$ denote the $L$ subpackets of the message $\mathrm{X}_k$, arranged in  a randomly chosen order. 
These orderings are generated independently for each message, and are known only to the user, not the servers.

The proposed scheme consists of three steps as described below. 

{\bf Step 1:} 
The user constructs $N$ vectors $\{\mathrm{v}_n\}_{n\in [N]}$, each of length $KL$ with entries from $\mathbbmss{F}_q$. 
Each vector \[{\mathrm{v}_n = [v_{1,1},\dots,v_{1,L},\dots,v_{K,1},\dots,v_{K,L}]}\] represents the coefficient vector corresponding to a linear combination of the $KL$ message subpackets \[\mathrm{X}_{1,1},\dots,\mathrm{X}_{1,L},\dots,\mathrm{X}_{K,1},\dots,\mathrm{X}_{K,L}\] with combination coefficients from $\mathbbmss{F}_q$. 
These vectors are generated using a randomized algorithm---based on the demand's index set $\mathrm{W}$---as described below.

The user randomly selects a pair $(i,j)$ for some ${0\leq i\leq K-D}$ and ${1\leq j\leq D}$, where the probability of each pair $(i,j)$ is equal to $P_{i,j}$. 
The probabilities $\{{P}_{i,j}\}$ are defined as follows. 

Let $j^{*}\in [D]$ be an arbitrary index such that \[\frac{f_{j^{*}}}{g_{j^{*}}}= \max_{j\in [D]} \frac{f_j}{g_j},\] where 
${\mathrm{f} = [f_1,\dots,f_D]^{\mathsf{T}}}$ and ${\mathrm{g} = [g_1,\dots,g_D]^{\mathsf{T}}}$ are defined as in~\eqref{eq:f} and \eqref{eq:g}, respectively.

The probabilities $P_{K-D,1},\dots,P_{K-D,D}$ are defined as
\begin{equation}\label{eq:PKD}
{P_{K-D,i}}=
\begin{cases}
      {1}/{g_{j^{*}}} & \text{if $i=j^{*}$,}\\
      0 & \text{otherwise,}
    \end{cases} 
\end{equation}
and for each $0\leq i\leq K-D-1$, the probabilities $P_{i,1},\dots,P_{i,D}$ are given by 
\begin{equation}\label{eq:PiPKD}
\begin{array}{lr}
\begin{bmatrix}
P_{i,1}\\
\vdots\\
P_{i,D}
\end{bmatrix}
= C_{K-D,i}\hspace{2pt}\mathrm{M}^{K-D-i}
\begin{bmatrix}
P_{K-D,1}\\
\vdots\\
P_{K-D,D}
\end{bmatrix},
\end{array}
\end{equation} where $\mathrm{M}$ is defined as in~\eqref{eq:M}. 

Given the selected pair $(i,j)$, 
the user randomly generates an \emph{$i$-sparse} vector $\mathrm{h}$ of length $K-D$ over $\mathbbmss{F}_q$ that has $i$ nonzero entries in randomly chosen positions, and a \emph{$j$-regular} invertible $D\times D$ matrix $\mathrm{G} = [\mathrm{g}^{\mathsf{T}}_1,\dots,\mathrm{g}^{\mathsf{T}}_D]^{\mathsf{T}}$ over $\mathbbmss{F}_q$ satisfying the following properties: 
(i) the vector $\mathrm{g}_1$ has $j$ nonzero entries in randomly chosen positions, and 
(ii) for each $2\leq m\leq D$, the positions of the nonzero entries in the vector $\mathrm{g}_m$ are a circular shift of those in the vector $\mathrm{g}_{m-1}$. 

Next, the user constructs the vectors $\{\mathrm{v}_n\}$ as follows: 

\begin{itemize}
\item If $i=0$, then $\mathrm{v}_{1}$ is an all-zero vector. 
Otherwise, if ${i\neq 0}$, $\mathrm{v}_{1}$ is the coefficient vector corresponding to the linear combination $\mathrm{Y}_{1}$ defined as
\[\mathrm{Y}_1 = \mathrm{h}\cdot [\mathrm{X}_{u_1,1},\dots,\mathrm{X}_{u_{K-D},1}]^{\mathsf{T}},\] 
where ${u_1,\dots,u_{K-D}}$ are the indices of $K-D$ interference messages, arranged in an arbitrary but fixed (e.g., ascending or descending) order; 
\item For each $1 \leq l \leq L$ and $1 \leq m \leq D$, the vector $\mathrm{v}_{(l-1)D +m+1}$ is the coefficient vector corresponding to the linear combination $\mathrm{Y}_{(l-1)D+m+1}$ defined as
\[\mathrm{Y}_{(l-1)D+m+1} := \mathrm{Y}_{1} + {\mathrm{g}_{m}} \cdot [\mathrm{X}_{w_{1},l}, \dots, \mathrm{X}_{w_{D},l}]^{\mathsf{T}},\] 
where $w_{1},\dots,w_{D}$ are the indices of the $D$ demand messages, arranged in an arbitrary but fixed order.
\end{itemize} 

Note that $\mathrm{Y}_1$ is a linear combination of the first subpackets of $i$ randomly chosen interference messages. 
Moreover, for each $1\leq m\in D$, 
$\mathrm{Y}_{m+1}-\mathrm{Y}_1$ is a linear combination of the first subpackets of $j$ randomly chosen demand messages, $\mathrm{Y}_{D+m+1}-\mathrm{Y}_1$ is a linear combination of the second subpackets of the same $j$ demand messages, and so forth. 

\vspace{0.125cm}
{\bf Step 2.} 
The user randomly selects a permutation ${\pi: [N]\rightarrow [N]}$, and for each $n\in [N]$, sends the vector $\mathrm{v}_{\pi(n)}$ as the query $\mathrm{Q}_{n}^{[\mathrm{W}]}$ to server $n$.
Each server $n$ then computes the corresponding linear combination $\mathrm{Y}_{\pi(n)}$ and sends it back to the user as the answer $\mathrm{A}^{[\mathrm{W}]}_{n}$.

\vspace{0.125cm}
{\bf Step 3.} The user computes
${\mathrm{Z}_{n-1} := \mathrm{Y}_{n}-\mathrm{Y}_1}$ for all ${2\leq n\leq N}$. 
By construction, $\mathrm{Z}_1,\dots,\mathrm{Z}_{N-1}$ form $N-1$ ($=DL$) linearly independent combinations of the $DL$ subpackets of the demand messages. 
Thus, the user can recover all subpackets of the demand messages by solving the resulting system of linear equations.  

\subsection{Proof of Validity of $\{P_{i,j}\}$}
In this section, we show that our choice of $\{P_{i,j}\}$ defines a valid probability distribution. 

By definition, $P_{i,j}\geq 0$ for all ${0\leq i\leq K-D}$ and ${1\leq j\leq D}$. 
It remains to show that  
\begin{equation}\label{eq:NewSumProb}
\sum_{i=0}^{K-D}\sum_{j=1}^{D} P_{i,j} = 1.
\end{equation}

Let 
\begin{equation}\label{eq:Pi}
\mathrm{P}_i := [P_{i,1},\dots,P_{i,D}]^{\mathsf{T}}   
\end{equation} for all $0\leq i\leq K-D$. 
Note that 
\begin{equation}\label{eq:PijSum}
\sum_{j=1}^{D} P_{i,j} = \mathbf{1}^{\mathsf{T}}\mathrm{P}_i
\end{equation}
for all ${0\leq i\leq K-D}$, where $\mathbf{1}$ is the all-one column-vector of length $D$.
Note also that
\begin{equation}\label{eq:FinalFinalCond}
 \mathrm{P}_{i} \stackrel{\footnotesize\text{\eqref{eq:PiPKD}}}{=} C_{K-D,i}\hspace{2pt}\mathrm{M}^{K-D-i}\mathrm{P}_{K-D} 
\end{equation} for all ${0\leq i\leq K-D-1}$. 
Thus, we have
\begin{align*}
\sum_{i=0}^{K-D}\sum_{j=1}^{D} P_{i,j} & \stackrel{\footnotesize\text{\eqref{eq:PijSum}}}{=} \sum_{i=0}^{K-D} \mathbf{1}^{\mathsf{T}}\mathrm{P}_i \\
& \stackrel{\footnotesize\text{\eqref{eq:FinalFinalCond}}}{=} \sum_{i=0}^{K-D} \mathbf{1}^{\mathsf{T}}C_{K-D,i}\hspace{2pt} \mathrm{M}^{K-D-i}\mathrm{P}_{K-D} \\
& = \mathbf{1}^{\mathsf{T}} (\mathrm{I}+\mathrm{M})^{K-D} \mathrm{P}_{K-D}\\
& \stackrel{\footnotesize\text{\eqref{eq:g}}}{=} \mathrm{g}^{\mathsf{T}} \mathrm{P}_{K-D}. 
\end{align*}
Rewriting in matrix form,~\eqref{eq:NewSumProb} can be expressed as
\begin{equation}\label{eq:FFFCond}
\mathrm{g}^{\mathsf{T}} \mathrm{P}_{K-D} = 1,    
\end{equation} which follows from our choice of $P_{K-D,1},\dots,P_{K-D,D}$ in~\eqref{eq:PKD}, since
\begin{align*}
\mathrm{g}^{\mathsf{T}} \mathrm{P}_{K-D} \stackrel{\footnotesize\text{\eqref{eq:PKD}}}{=} g_{j^{*}} P_{K-D,j^{*}} = 1.
\end{align*} 

\subsection{Proofs of Recoverability and Privacy}\label{sec:OptimalProb}
The proof of recoverability follows from the construction of the queries and answers, and is therefore omitted. 
To prove that the privacy condition is satisfied, we proceed as follows. 

Fix an arbitrary $n\in [N]$.
Let ${\mathrm{v}=[v_{1,1},\dots,v_{K,L}]}$ be the (query) vector sent by the user to server $n$, and let \[{\mathrm{T} = \{(k,l)\in [K]\times [L]: v_{k,l}\neq 0\}}\] be the index set of message subpackets contributing to the linear combination corresponding to the vector $\mathrm{v}$.

Let $\mathbf{v}$ and $\mathbf{T}$ denote the random variables corresponding to $\mathrm{v}$ and $\mathrm{T}$, respectively. 

To satisfy the privacy condition, ${\mathbb{P}(\mathbf{W}=\mathrm{W}|\mathbf{v}=\mathrm{v})}$ must be identical for all  ${\mathrm{W}\in \mathbbmss{W}}$. 
Since $\mathbf{W}$ is distributed uniformly over $\mathbbmss{W}$, the privacy condition is met so long as  
${\mathbb{P}(\mathbf{v}=\mathrm{v}|\mathbf{W}=\mathrm{W})}$ is the same for all ${\mathrm{W}\in \mathbbmss{W}}$.  

Since $\mathrm{T}$ is a deterministic function of $\mathrm{v}$, we can write
\begin{align*}
&\mathbb{P}(\mathbf{v}=\mathrm{v}|\mathbf{W}=\mathrm{W}) = \mathbb{P}( \mathbf{v}=\mathrm{v},\mathbf{T} = \mathrm{T}|\mathbf{W}=\mathrm{W})\\
& = \mathbb{P}(\mathbf{T}=\mathrm{T} |\mathbf{W}=\mathrm{W})\times \mathbb{P}(\mathbf{v}=\mathrm{v}|\mathbf{W}=\mathrm{W},\mathbf{T}=\mathrm{T}).
\end{align*}
Note that ${\mathbb{P}(\mathbf{v} = \mathrm{v}|\mathbf{W}=\mathrm{W},\mathbf{T}=\mathrm{T})}$ does not depend on $\mathrm{W}$, since the values of nonzero entries of $\mathrm{v}$ are chosen independently of $\mathrm{W}$. 
Thus, the privacy condition is satisfied so long as ${\mathbb{P}(\mathbf{T}=\mathrm{T}|\mathbf{W}=\mathrm{W})}$ is the same for all  ${\mathrm{W}\in \mathbbmss{W}}$. 

Recall that, by construction, each message either has no subpackets contributing to the linear combination corresponding to $\mathrm{v}$ or one subpacket contributing.
Let \[{\mathrm{S} = \{k\in [K]: (k,l)\in \mathrm{T} \text{ for some } l\in [L]\}}\] be the index set of messages with one subpacket contributing, and let $\mathbf{S}$ denote the random variable corresponding to $\mathrm{S}$.  

Since the subpackets of each message are indexed independently of $\mathrm{W}$, 
${\mathbb{P}(\mathbf{T}=\mathrm{T}|\mathbf{W}=\mathrm{W})}$ is the same for all ${\mathrm{W}\in \mathbbmss{W}}$ so long as ${\mathbb{P}(\mathbf{S}=\mathrm{S}|\mathbf{W}=\mathrm{W})}$ is the same for all ${\mathrm{W}\in \mathbbmss{W}}$. 

Let $S = |\mathrm{S}|$. 
Fix an arbitrary $\mathrm{W}\in \mathbbmss{W}$. 
Let $I = |\mathrm{S}\setminus \mathrm{W}|$ and $J = |\mathrm{S}\cap \mathrm{W}|$. 
By construction, we have ${0\leq I\leq \min\{S,K-D\}}$ and ${0\leq J\leq \min\{S,D\}}$.
 
If $J=0$, then $\mathrm{S}$ contains $I$ interference message indices and no demand message indices. 
This implies that the value of $i$ selected by the scheme is equal to $I$, and $\mathrm{S}$ corresponds to the vector $\mathrm{v}_1$ generated by the scheme, which corresponds to a linear combination of the first subpackets of $I$ randomly chosen interference messages. 
Since any given $I$-subset of interference message indices is used for construction with probability $1/C_{K-D,I}$, and $\mathrm{v}_1$ is constructed independent of the value of $j$ selected by the scheme, we have 
\begin{equation}\label{eq:PTempty}
\mathbb{P}(\mathbf{S}=\mathrm{S}|\mathbf{W}=\mathrm{W}) = \frac{1}{N}\sum_{j=1}^{D} \frac{1}{C_{K-D,I}} \times P_{I,j},
\end{equation} where $1/N$ is the probability that the user selects $\mathrm{v}_1$ from the set of $N$ vectors $\mathrm{v}_1,\dots,\mathrm{v}_N$ to send to server $n$. 

If $J\neq 0$, then $\mathrm{S}$ contains $I$ interference message indices and $J$ demand message indices. 
That is, the values of $i$ and $j$ selected by the scheme are equal to $I$ and $J$, respectively. 
We consider the cases $1\leq J\leq D-1$ and $J=D$ separately. 

If $1\leq J\leq D-1$, then $\mathrm{S}$ corresponds to the $L$ vectors $\mathrm{v}_{m+1},\mathrm{v}_{D+m+1},\dots,\mathrm{v}_{(L-1)D+m+1}$ generated by the scheme for some $m\in [D]$, which correspond to linear combinations of different subpackets of the same $J$ demand messages. 
Since $\mathrm{v}_2$ involves $J$ randomly chosen demand message indices, and the demand message indices involved in $\mathrm{v}_3,\dots,\mathrm{v}_D$ are obtained by circularly shifting those involved in $\mathrm{v}_2$, 
the probability of any given $J$-subset of demand message indices being used for construction is given by $D/C_{D,J}$. 
Thus, we have 
\begin{equation}\label{eq:PTnonempty1}
\mathbb{P}(\mathbf{S}=\mathrm{S}|\mathbf{W}=\mathrm{W}) = \frac{L}{N}\times \frac{1}{C_{K-D,I}}\times \frac{D}{C_{D,J}}\times P_{I,J}, 
\end{equation} 
where $L/N$ is the probability that the user sends one of the $L$ vectors $\mathrm{v}_{m+1},\mathrm{v}_{D+m+1},\dots,\mathrm{v}_{(L-1)D+m+1}$ to server $n$. 

If $J = D$, then $\mathrm{S}$ corresponds to all $N-1$ ($=DL$) vectors $\mathrm{v}_{2},\dots,\mathrm{v}_{N}$ generated by the scheme, which correspond to linear combinations of different subpackets of all $D$ demand messages. 
Thus, we have
\begin{equation}\label{eq:PTnonempty2}
\mathbb{P}(\mathbf{S}=\mathrm{S}|\mathbf{W}=\mathrm{W}) = \frac{DL}{N}\times \frac{1}{C_{K-D,I}}\times  P_{I,D}, 
\end{equation} 
where $DL/N$ is the probability that the user sends one of the $DL$ vectors $\mathrm{v}_{2},\dots,\mathrm{v}_{N}$ to server $n$.

By combining~\eqref{eq:PTnonempty1} and~\eqref{eq:PTnonempty2}, we have, for all $J\neq 0$,
\begin{equation}\label{eq:PTnonempty}
\mathbb{P}(\mathbf{S}=\mathrm{S}|\mathbf{W}=\mathrm{W}) = \frac{DL}{N}\times \frac{1}{C_{K-D,I}}\times \frac{1}{C_{D,J}}\times P_{I,J}.
\end{equation}

For simplicity, let \[\alpha_i := \frac{1}{C_{K-D,i}}\quad \text{and} \quad \beta_j := \frac{DL}{C_{D,j}},\] for $0\leq i\leq K-D$ and $1\leq j\leq D$, respectively.

Recall that the probabilities $\{P_{i,j}\}$ must be chosen such that ${\mathbb{P}(\mathbf{S}=\mathrm{S}|\mathbf{W}=\mathrm{W})}$ is the same for all ${\mathrm{W}\in \mathbbmss{W}}$. 
 
First, suppose that $\mathrm{S}= \emptyset$ (i.e., $S = 0$). 
Since, for any ${\mathrm{W}\in \mathbbmss{W}}$,  ${|\mathrm{S}\setminus \mathrm{W}|=0}$ and ${|\mathrm{S}\cap \mathrm{W}|=0}$, it follows from~\eqref{eq:PTempty} that ${\mathbb{P}(\mathbf{S}=\emptyset|\mathbf{W}=\mathrm{W})} = \frac{1}{N}\sum_{j=1}^{D}  P_{0,j}$, which is the same for all ${\mathrm{W}\in \mathbbmss{W}}$, regardless of the choice of $\{P_{i,j}\}$.  

Next, suppose that $\mathrm{S}\neq \emptyset$ (i.e., ${1\leq S\leq K}$). 
Note that, for any ${\mathrm{W}\in \mathbbmss{W}}$, we have 
${0\leq |\mathrm{S}\setminus \mathrm{W}|\leq \min\{S,K-D\}}$ and
${0\leq |\mathrm{S}\cap \mathrm{W}|\leq \min\{S,D\}}$. 

For any ${0\leq T\leq\min\{S,D\}}$, let $\mathbbmss{W}_{T}$ be the collection of all ${\mathrm{W}\in \mathbbmss{W}}$ such that ${|\mathrm{S}\cap \mathrm{W}|=T}$. 

We consider the following two cases separately: 
(i)~${1\leq S\leq K-D}$, and 
(ii) ${K-D+1\leq S\leq K}$. 

First, consider the case (i). 
In this case, we have ${\mathbbmss{W}_T\neq \emptyset}$ for all ${0\leq T\leq \min\{S,D\}}$.

For all ${\mathrm{W}\in \mathbbmss{W}_0}$, we have ${|\mathrm{S}\setminus \mathrm{W}|=S}$ and ${|\mathrm{S}\cap \mathrm{W}| = 0}$. 
Thus, 
\begin{align}\label{eq:=0}
& \mathbb{P}(\mathbf{S} =\mathrm{S}|\mathbf{W}=\mathrm{W}) \stackrel{\footnotesize\text{\eqref{eq:PTempty}}}{=} \frac{1}{N}\sum_{j=1}^{D} \alpha_S P_{S,j}.  
\end{align} 

For any ${1\leq T\leq \min\{S,D\}}$, we have ${|\mathrm{S}\setminus \mathrm{W}|=S-T}$ and ${|\mathrm{S}\cap \mathrm{W}|=T}$ for all ${\mathrm{W}\in \mathbbmss{W}_T}$. 
Thus, 
\begin{align}\label{eq:neq0}
& \mathbb{P}(\mathbf{S}=\mathrm{S}|\mathbf{W}=\mathrm{W})\stackrel{\footnotesize\text{\eqref{eq:PTnonempty}}}{=} \frac{1}{N} \alpha_{S-T}\beta_{T} P_{S-T,T}. 
\end{align} 
By~\eqref{eq:=0} and~\eqref{eq:neq0}, for any $\mathrm{S}$ such that ${1\leq S\leq K-D}$, ${\mathbb{P}(\mathbf{S}=\mathrm{S}|\mathbf{W}=\mathrm{W})}$ is the same for all ${\mathrm{W}\in \mathbbmss{W}}$ so long as
\begin{equation}\label{eq:orgCond11}
\sum_{j=1}^{D} \alpha_S P_{S,j} = \alpha_{S-1} \beta_{1} P_{S-1,1}
\end{equation} 
and
\begin{equation}\label{eq:orgCond12}
\alpha_{S-T} \beta_{J} P_{S-T,T} = \alpha_{S-T-1} \beta_{T+1} P_{S-T-1,T+1} 
\end{equation}
for all ${1\leq T\leq \min\{S,D\}-1}$. 

Next, consider the case (ii). 
In this case, we have ${\mathbbmss{W}_T=\emptyset}$ for all ${0\leq T\leq S-K+D-1}$, and ${\mathbbmss{W}_T\neq \emptyset}$ for all ${S-K+D\leq T\leq \min\{S,D\}}$. 

For any ${S-K+D\leq T\leq \min\{S,D\}}$, we have ${|\mathrm{S}\setminus \mathrm{W}|=S-T}$ and ${|\mathrm{S}\cap \mathrm{W}|=T}$ for all ${\mathrm{W}\in \mathbbmss{W}_T}$. 
Thus,
\begin{equation}\label{eq:PCaseii}
\mathbb{P}(\mathbf{S}=\mathrm{S}|\mathbf{W}=\mathrm{W}) \stackrel{\footnotesize\text{\eqref{eq:PTnonempty}}}{=} \frac{1}{N}\alpha_{S-T}\beta_{T} P_{S-T,T}.
\end{equation}
By~\eqref{eq:PCaseii}, for any $\mathrm{S}$ such that ${K-D+1\leq S\leq K}$, ${\mathbb{P}(\mathbf{S}=\mathrm{S}|\mathbf{W}=\mathrm{W})}$ is the same for all ${\mathrm{W}\in \mathbbmss{W}}$ so long as
\begin{equation}\label{eq:orgCond2}
\alpha_{S-T} \beta_{T} P_{S-T,T} = \alpha_{S-T-1} \beta_{T+1} P_{S-T-1,T+1}
\end{equation} for all ${S-K+D\leq T\leq \min\{S,D\}-1}$. 

By these arguments, the privacy condition is satisfied so long as the conditions~\eqref{eq:orgCond11}, \eqref{eq:orgCond12} and~\eqref{eq:orgCond2} are met.  
By a series of simple change of variables, it can be shown that these conditions are met simultaneously so long as the probabilities $\{P_{i,j}\}$ satisfy the following two conditions: 
\begin{equation}\label{eq:FinalCond1}
\sum_{j=1}^{D}\alpha_i P_{i,j} = \alpha_{i-1}\beta_1 P_{i-1,1}    
\end{equation} for all ${1\leq i\leq K-D}$, and
\begin{equation}\label{eq:FinalCond2}
\alpha_i \beta_j P_{i,j} = \alpha_{i-1} \beta_{j+1} P_{i-1,j+1}
\end{equation} for all ${1\leq i\leq K-D}$ and ${1\leq j\leq D-1}$. 

Rewriting in matrix form, it follows that both conditions~\eqref{eq:FinalCond1} and~\eqref{eq:FinalCond2}, and hence the privacy condition, are satisfied so long as 
\begin{equation}\label{eq:FinalCond}
\alpha_{i-1}\mathrm{P}_{i-1} = \alpha_i \mathrm{M}\mathrm{P}_i   
\end{equation} for all ${1\leq i\leq K-D}$, or equivalently, 
\begin{equation*}
 \mathrm{P}_{i} = C_{K-D,i}\hspace{2pt}\mathrm{M}^{K-D-i}\mathrm{P}_{K-D} 
\end{equation*} for all ${0\leq i\leq K-D-1}$, 
which coincides with the choice of probabilities $P_{i,1},\dots,P_{i,D}$ for ${0\leq i\leq K-D-1}$ as defined in~\eqref{eq:PiPKD}. 
This completes the proof of privacy. 

\subsection{Proof of Achievable Rate}\label{subsec:RateCalc}
It remains to show that the proposed scheme achieves the rate $R$ defined in~\eqref{eq:R}. 

Recall that the rate is defined as the ratio of $H(\mathbf{X}_{\mathbf{W}})$ to ${\sum_{n=1}^{N} H(\mathbf{A}^{[\mathbf{W}]}_n|\mathbf{Q}^{[\mathbf{W}]}_n)}$.   
Since $\mathbf{W}$ and $\mathbf{X}_1,\dots,\mathbf{X}_K$ are independent, and 
$\mathbf{X}_1,\dots,\mathbf{X}_K$ are independent and uniformly distributed over $\mathbbmss{F}_q^m$, it follows that $H(\mathbf{X}_{\mathbf{W}}) = DB$, where $B := m\log_2 q$ is the entropy of a uniform random variable over $\mathbbmss{F}_q^m$, measured in bits. 
It then remains to compute ${\sum_{n=1}^{N} H(\mathbf{A}^{[\mathbf{W}]}_n|\mathbf{Q}^{[\mathbf{W}]}_n)}$. 

Fix an arbitrary $n\in [N]$. 
Since $\mathbf{W}$ and $\mathbf{X}_1,\dots,\mathbf{X}_K$ are independent, and  $\mathbf{A}^{[\mathbf{W}]}_n$ is a function of $\mathbf{Q}^{[\mathbf{W}]}_n$ and $\mathbf{X}_1,\dots,\mathbf{X}_K$, $\mathbf{W}\leftrightarrow \mathbf{Q}^{[\mathbf{W}]}_n\leftrightarrow \mathbf{A}^{[\mathbf{W}]}_n$ forms a Markov chain. 
Thus, we have 
\[H(\mathbf{A}^{[\mathbf{W}]}_n|\mathbf{Q}^{[\mathbf{W}]}_n) = \sum_{\mathrm{W}\in \mathbbmss{W}} \mathbb{P}(\mathbf{W} = \mathrm{W})H(\mathbf{A}_n^{[\mathrm{W}]}|\mathbf{Q}_n^{[\mathrm{W}]}).\] 
Note that $H(\mathbf{A}_n^{[\mathrm{W}]}|\mathbf{Q}_n^{[\mathrm{W}]})$ is given by \[\sum \mathbb{P}(\mathbf{Q}^{[\mathrm{W}]}_n = \mathrm{Q}^{[\mathrm{W}]}_n) H(\mathbf{A}^{[\mathrm{W}]}_n|\mathbf{Q}^{[\mathrm{W}]}_n = \mathrm{Q}^{[\mathrm{W}]}_n),\] where the sum is over all possible realizations $\mathrm{Q}^{[\mathrm{W}]}_n$ of $\mathbf{Q}^{[\mathrm{W}]}_n$.

Fix an arbitrary $\mathrm{W}\in \mathbbmss{W}$. 
By construction, 
either 
(i) server $n$ receives an all-zero vector as the query from the user (i.e., $\mathrm{Q}^{[\mathrm{W}]}_n=0$) and hence, server $n$ does not send any answer back to the user, 
or 
(ii) server $n$ receives a nonzero vector as the query from the user (i.e., $\mathrm{Q}^{[\mathrm{W}]}_n\neq 0$), and the answer of server $n$ to the user's query is a nontrivial linear combination of the message subpackets.
In the case (i), ${H(\mathbf{A}^{[\mathrm{W}]}_n|\mathbf{Q}^{[\mathrm{W}]}_n=\mathrm{Q}^{[\mathrm{W}]}_n)=0}$ since server $n$ does not send any answer back to the user, 
whereas in the case (ii), ${H(\mathbf{A}^{[\mathrm{W}]}_n|\mathbf{Q}^{[\mathrm{W}]}_n = \mathrm{Q}^{[\mathrm{W}]}_n)= B/L}$, since each message consists of $L$ subpackets, identically and uniformly distributed over $\mathbbmss{F}_q$, and any nonzero linear combination of them is uniformly distributed over $\mathbbmss{F}_q$.
Thus,
\begin{align*}
H(\mathbf{A}^{[\mathrm{W}]}_n|\mathbf{Q}^{[\mathrm{W}]}_n)=\left(1-\mathbb{P}\left(\mathbf{Q}^{[\mathrm{W}]}_n = 0\right)\right)B/L.
\end{align*}

Since ${\mathbf{Q}^{[\mathrm{W}]}_n = 0}$ with probability $\frac{1}{N}\sum_{j=1}^{D} P_{0,j}$, we have
\begin{align*}
H(\mathbf{A}^{[\mathrm{W}]}_n|\mathbf{Q}^{[\mathrm{W}]}_n) = \left(1- \frac{1}{N}\sum_{j=1}^{D} P_{0,j}\right)B/L.
\end{align*}
Note that $H(\mathbf{A}^{[\mathrm{W}]}_n|\mathbf{Q}^{[\mathrm{W}]}_n)$ does not depend on $\mathrm{W}$.  
Thus,
\[
H(\mathbf{A}^{[\mathbf{W}]}_n|\mathbf{Q}^{[\mathbf{W}]}_n) = \left(1- \frac{1}{N}\sum_{j=1}^{D} P_{0,j}\right)B/L.
\]
Since $H(\mathbf{A}^{[\mathbf{W}]}_n|\mathbf{Q}^{[\mathbf{W}]}_n)$ does not depend on $n$, we have 
\[H(\mathbf{A}^{[\mathbf{W}]}_1|\mathbf{Q}^{[\mathbf{W}]}_1) = \dots = H(\mathbf{A}^{[\mathbf{W}]}_N|\mathbf{Q}^{[\mathbf{W}]}_N).\]
Thus, 
\begin{equation}\label{eq:DC}
\sum_{n=1}^{N} H(\mathbf{A}^{[\mathbf{W}]}_n|\mathbf{Q}^{[\mathbf{W}]}_n) = \left(N-\sum_{j=1}^{D} P_{0,j}\right)B/L,    
\end{equation}
which implies that the rate of the scheme is given by
\begin{equation}\label{eq:Rate}
\frac{DL}{N-\sum_{j=1}^{D} P_{0,j}}.
\end{equation}

Since
\begin{equation}\label{eq:ObjMatrix}
\sum_{j=1}^{D} P_{0,j} = \mathbf{1}^{\mathsf{T}}\mathrm{P}_{0} \stackrel{\footnotesize\text{\eqref{eq:FinalFinalCond}}}{=} \mathbf{1}^{\mathsf{T}}\mathrm{M}^{K-D}\mathrm{P}_{K-D} \stackrel{\footnotesize\text{\eqref{eq:f}}}{=} \mathrm{f}^{\mathsf{T}} \mathrm{P}_{K-D},
\end{equation} 
we can rewrite~\eqref{eq:Rate} as 
\begin{equation}\label{eq:FinalRate}
\frac{DL}{N-\mathrm{f}^{\mathsf{T}} \mathrm{P}_{K-D}}.
\end{equation}
Note that
\begin{equation}\label{eq:fjgj}
\mathrm{f}^{\mathsf{T}} \mathrm{P}_{K-D} = f_{j^{*}}P_{K-D,j^{*}} \stackrel{\footnotesize\text{\eqref{eq:PKD}}}{=} f_{j^{*}}/g_{j^{*}}
\end{equation} 
by our choice of $P_{K-D,1},\dots,P_{K-D,D}$ in~\eqref{eq:PKD}.
Substituting~\eqref{eq:fjgj} into~\eqref{eq:FinalRate}, it then follows that the rate is given by 
\[\frac{DL}{N-f_{j^{*}}/g_{j^{*}}},\] which is the same as $R$ defined in~\eqref{eq:R}.
 
\subsection{Proof of Optimality of $\{P_{i,j}\}$}
In this section, we show the optimality of our choice of probabilities $\{P_{i,j}\}$ in the proposed scheme.  

Recall that $\{P_{i,j}\}$ must define a valid probability distribution, i.e., it must hold that 
\begin{equation}\label{eq:PijNN}
P_{i,j}\geq 0
\end{equation}
for all $0\leq i\leq K-D$ and $1\leq j\leq D$, and 
\begin{equation}\label{eq:SumOpt}
\sum_{i=0}^{K-D}\sum_{j=1}^{D} P_{i,j} = 1.
\end{equation}
In addition, $\{P_{i,j}\}$ must be carefully chosen to satisfy the privacy condition, i.e., 
\begin{equation}\label{eq:CondOpt}
\begin{array}{lr}
\begin{bmatrix}
P_{i,1}\\
\vdots\\
P_{i,D}
\end{bmatrix}
= C_{K-D,i}\hspace{2pt}\mathrm{M}^{K-D-i}
\begin{bmatrix}
P_{K-D,1}\\
\vdots\\
P_{K-D,D}
\end{bmatrix},
\end{array}
\end{equation} for all $0\leq i\leq K-D-1$, where $\mathrm{M}$ is defined in~\eqref{eq:M}.  

Let $\mathrm{P}_i = [P_{i,1},\dots,P_{i,D}]$ for all $0\leq i\leq K-D$. 
Since, by~\eqref{eq:CondOpt}, $\mathrm{P}_i$ for $0\leq i\leq K-D-1$ are determined given $\mathrm{P}_{K-D}$,~\eqref{eq:PijNN} and~\eqref{eq:SumOpt} can be written in terms of $\mathrm{P}_{K-D}$ as
\begin{equation}\label{eq:PKDNN}
\mathrm{P}_{K-D}\geq \mathbf{0},
\end{equation} and
\begin{equation}\label{eq:FFFCondOpt}
\mathrm{g}^{\mathsf{T}} \mathrm{P}_{K-D} = 1,    
\end{equation} where $\mathbf{0}$ denotes the all-zero column-vector of length $D$, and $\mathrm{g}$ is defined in~\eqref{eq:g}.  

Recall that the rate of the scheme is given by 
\begin{equation*}
\frac{DL}{N-\mathrm{f}^{\mathsf{T}}\mathrm{P}_{K-D}},
\end{equation*} 
where $\mathrm{f}$ is defined in~\eqref{eq:f}. 
To maximize the rate, we need to maximize $\mathrm{f}^{\mathsf{T}}\mathrm{P}_{K-D}$, subject to the constraints in~\eqref{eq:PKDNN} and~\eqref{eq:FFFCondOpt}. 
Thus, the optimal choice of the vector $\mathrm{P}_{K-D}$ is obtained by solving the following linear program (LP): 
\begin{align*}
\mathrm{max} & \quad \mathrm{f}^{\mathsf{T}}\mathrm{P}_{K-D}\\
\mathrm{s.t.} & \quad \mathrm{g}^{\mathsf{T}} \mathrm{P}_{K-D} = 1\nonumber\\
& \quad \mathrm{P}_{K-D}\geq \mathbf{0}\nonumber
\end{align*} 
which can be rewritten in component form as
\begin{align*}
\mathrm{max} & \quad \sum_{j=1}^{D} f_j P_{K-D,j}\\
\mathrm{s.t.} & \quad \sum_{j=1}^{D} g_j P_{K-D,j} =  1\nonumber\\
& \quad P_{K-D,j}\geq 0 \quad  \forall 1\leq j\leq D\nonumber
\end{align*} 
where $f_j$ and $g_j$ are the $j$th entries of $\mathrm{f}$ and $\mathrm{g}$, respectively.

It is easy to verify that the optimal solution to this LP problem is given by ${P_{K-D,j^{*}} = {1}/{g_{j^{*}}}}$ and 
${P_{K-D,j} = 0}$ for all ${j\in [D]\setminus \{j^{*}\}}$, 
where $j^{*}\in [D]$ is an arbitrary index such that $f_{j^{*}}/g_{j^{*}}\geq f_j/g_j$ for all $j\in [D]$. 
This matches the probabilities $P_{K-D,1},\dots,P_{K-D,D}$ in~\eqref{eq:PKD}. 
Moreover, the probabilities $P_{i,j}$ for $0\leq i\leq K-D-1$ and $1\leq j\leq D$ are determined using~\eqref{eq:CondOpt}, aligning with those in~\eqref{eq:PiPKD}.  

\section{Proof of Theorem~\ref{thm:2}}
To prove Theorem~\ref{thm:2}, we show that when $D\mid K$,  
the rate $R$ defined in~\eqref{eq:R} is equal to
\begin{equation}\label{eq:SimCap1}
\frac{1-1/N}{1-1/N^{K/D}},    
\end{equation}
where $N=DL+1$ for any $L\geq 1$.

It is easy to verify that~\eqref{eq:SimCap1} can be written as 
\begin{equation}\label{eq:SimCap2}
\frac{DL}{N-(DL+1)^{1-K/D}}.
\end{equation}
Comparing~\eqref{eq:Rate} and~\eqref{eq:SimCap2}, we need to show that 
\begin{equation}\label{eq:ToShow}
\sum_{j=1}^{D} P_{0,j} = (DL+1)^{1-K/D}.  \end{equation}

By combining~\eqref{eq:FinalCond1} and~\eqref{eq:FinalCond2}, it can be shown that
\begin{equation}\label{eq:Rec1}
\alpha_i P_{i,j} = \sum_{h=1}^{D} \frac{1}{\beta_h} \alpha_{i+h} P_{i+h,j} 
\end{equation} for all ${0\leq i\leq K-2D}$ and ${1\leq j\leq D}$. 

Defining $O_{i,j}:= \alpha_{K-D-i} P_{K-D-i,j}$ for all $0\leq i\leq K-D$ and $1\leq j\leq D$, we can rewrite~\eqref{eq:Rec1} as 
\begin{equation}\label{eq:Rec2}
O_{i,j} = \sum_{h=1}^{D} \frac{1}{\beta_h} O_{i-h,j} \end{equation} for all ${D\leq i\leq K-D}$ and ${1\leq j\leq D}$. 
It is easy to see that~\eqref{eq:Rec2} represents $D$ linear recurrence relations of order $D$ in variable $i$---one recurrence for each ${1\leq j\leq D}$. 
Note that the $D$ initial values for the $j$th recurrence relation are $O_{0,j},\dots,O_{D-1,j}$, and these $D$ recurrence relations may have different initial values.  

The characteristic equation for each of these recurrence relations is given by 
${\lambda^D - \sum_{h=1}^{D}\frac{1}{\beta_h}\lambda^{D-h} = 0}$,
which is equivalent to
${DL\lambda^D - \sum_{h=1}^{D}\binom{D}{h}\lambda^{D-h} = 0}$.  
By the binomial theorem, ${\sum_{h=1}^{D} \binom{D}{h}\lambda^{D-h} =  (1+\lambda)^D-\lambda^D}$, and 
hence, the characteristic equation can be written as
${(DL+1)\lambda^D - (1+\lambda)^{D} = 0}$,  
which is a polynomial equation of degree $D$. 
Let $\lambda_1,\dots,\lambda_D$ be the roots of the characteristic equation. 
Note that 
\begin{equation}\label{eq:CE3}
(DL+1)\lambda^D_h = (1+\lambda_h)^{D}  
\end{equation} for all ${1\leq h\leq D}$. 
Then, the solution to the $j$th recurrence relation is given by 
\begin{equation}\label{eq:Qij}
O_{i,j} = \sum_{h=1}^{D} c_{j,h}\lambda^{i}_h    
\end{equation} 
for all ${0\leq i\leq K-D}$, 
where the coefficients ${c_{j,1},\dots,c_{j,D}}$ are determined by the initial values $O_{0,j},\dots,O_{D-1,j}$. 

Since $O_{i,j} = \alpha_{K-D-i} P_{K-D-i,j}$ by definition, it readily follows from~\eqref{eq:Qij} that 
\begin{equation*}
\alpha_i P_{i,j} = \sum_{h=1}^{D} c_{j,h}\lambda^{K-D-i}_h,  
\end{equation*} or equivalently, 
\begin{equation}\label{eq:Pij}
P_{i,j} = C_{K-D,i}\sum_{h=1}^{D} c_{j,h}\lambda^{K-D-i}_h  
\end{equation} 
for all ${0\leq i\leq K-D}$. 

By substituting~\eqref{eq:Pij} in~\eqref{eq:NewSumProb}, we can rewrite~\eqref{eq:NewSumProb} as
\begin{equation}\label{eq:New2SumProb}
\sum_{i=0}^{K-D}\sum_{j=1}^{D} C_{K-D,i} \sum_{h=1}^{D} c_{j,h}\lambda^{K-D-i}_{h} = 1. 
\end{equation} 
The LHS of~\eqref{eq:New2SumProb} can be rewritten as
\begin{align}
& \sum_{i=0}^{K-D}\sum_{j=1}^{D} C_{K-D,i} \sum_{h=1}^{D} c_{j,h}\lambda^{K-D-i}_{h} \nonumber\\
& \quad = \sum_{j=1}^{D} \sum_{h=1}^{D} c_{j,h}\sum_{i=0}^{K-D}C_{K-D,i}\lambda^{K-D-i}_{h} \label{eq:Int1}\\   
& \quad  = \sum_{j=1}^{D} \sum_{h=1}^{D} c_{j,h}\sum_{i=0}^{K-D}C_{K-D,i}\lambda^{i}_{h} \label{eq:Int2}\\
& \quad = \sum_{j=1}^{D} \sum_{h=1}^{D} c_{j,h} (1+\lambda_h)^{K-D},  \label{eq:Int3}  
\end{align} where~\eqref{eq:Int1} follows from changing the order of the summations;
\eqref{eq:Int2} follows from a simple change of variable and the symmetry of the binomial coefficients $C_{K-D,i}$;
and~\eqref{eq:Int3} follows from the binomial theorem. 

By using~\eqref{eq:Int3}, we can rewrite~\eqref{eq:New2SumProb} as
\begin{equation}\label{eq:New3SumProb}
\sum_{j=1}^{D} \sum_{h=1}^{D} c_{j,h} (1+\lambda_h)^{K-D} = 1.
\end{equation}
Raising both sides of~\eqref{eq:CE3} to the power of $K/D-1$, which is a non-negative integer, yields
\begin{equation}\label{eq:CE4}
(DL+1)^{K/D-1}\lambda^{K-D}_h = (1+\lambda_h)^{K-D}    
\end{equation} for all ${1\leq h\leq D}$. 
By combining~\eqref{eq:New3SumProb} and~\eqref{eq:CE4}, we have 
\begin{equation}\label{eq:New4SumProb}
\sum_{j=1}^{D} \sum_{h=1}^{D} c_{j,h} \lambda_h^{K-D} = (DL+1)^{1-K/D},
\end{equation} which directly implies~\eqref{eq:ToShow}. 
Specifically, we have
\begin{align}
\sum_{j=1}^{D} P_{0,j} & = \sum_{j=1}^{D} \sum_{h=1}^{D} c_{j,h} \lambda^{K-D}_h \label{eq:Last1}\\
& = (DL+1)^{1-K/D},\label{eq:Last2}
\end{align} 
where~\eqref{eq:Last1} follows from substituting $P_{0,j}$ using~\eqref{eq:Pij},
and~\eqref{eq:Last2} follows from~\eqref{eq:New4SumProb}. 

\section{An Illustrative Example}
In this section, we provide an example to illustrate the proposed scheme.

Consider a scenario in which $N=5$ servers store ${K=4}$ messages $\mathrm{X}_1,\mathrm{X}_2,\mathrm{X}_3,\mathrm{X}_4\in \mathbbmss{F}_q^m$ for a prime power $q\geq 3$ and an even integer $m\geq 2$, and a user wants to retrieve ${D=2}$ of these messages. 
Note that $L = (N-1)/D = 2$.

For simplicity, we denote the two demand message indices by $a$ and $b$ and the two interference message indices by $c$ and $d$. 
Additionally, each message, $\mathrm{X}_a$, $\mathrm{X}_b$, $\mathrm{X}_c$, and $\mathrm{X}_d$, is divided into $L=2$ subpackets, labeled independently and randomly as $\mathrm{X}_{a,1},\mathrm{X}_{a,2}$, $\mathrm{X}_{b,1},\mathrm{X}_{b,2}$, $\mathrm{X}_{c,1},\mathrm{X}_{c,2}$, and $\mathrm{X}_{d,1},\mathrm{X}_{d,2}$.

First, the user randomly selects a pair $(i,j)$ for some $0\leq i\leq 2$ and $1\leq j\leq 2$, where the probability of each pair $(i,j)$ is equal to $P_{i,j}$, defined as follows: 
${P_{0,1} = 2/15}$, ${P_{0,2} = 1/15}$, ${P_{1,1} = P_{1,2} = P_{2,1} = 4/15}$, and ${P_{2,2} =0}$. 

For this example, suppose the user selects the pair ${(i,j) = (1,1)}$, with probability $P_{1,1}=4/15$. 

Given the value of $i$, the user then constructs a random $i$-sparse vector $ [\boldsymbol{h}_1,\boldsymbol{h}_2]$ of length $2$, where the random variables $\boldsymbol{h}_1$ and $\boldsymbol{h}_2$ are defined as follows: 
\begin{itemize}
\item If $i=0$, then $\boldsymbol{h}_1=\boldsymbol{h}_2=0$. 
\item If $i=1$, then either 
(i) $\boldsymbol{h}_1$ is uniformly distributed over $\mathbbmss{F}_q^{\times}:=\mathbbmss{F}_q\setminus \{0\}$ and $\boldsymbol{h}_2=0$, or 
(ii) $\boldsymbol{h}_2$ is uniformly distributed over $\mathbbmss{F}_q^{\times}$ and $\boldsymbol{h}_1=0$, with probability $1/2$ for each case. 
\item If $i=2$, then $\boldsymbol{h}_1$ and $\boldsymbol{h}_2$ are i.i.d.~uniform over $\mathbbmss{F}_q^{\times}$.
\end{itemize}

Additionally, given the value of $j$, the user constructs a random $j$-regular invertible $2\times 2$ matrix 
\[\begin{bmatrix} \boldsymbol{g}_{1} & \boldsymbol{g}_{2}\\ \boldsymbol{g}_{3} & \boldsymbol{g}_{4}\end{bmatrix},\] 
where the random variables $\boldsymbol{g}_1,\dots,\boldsymbol{g}_4$ are defined as follows: 
\begin{itemize}
\item If $j=1$, then either 
(i) $\boldsymbol{g}_1$ and $\boldsymbol{g}_4$ are i.i.d.~uniform over $\mathbbmss{F}_q^{\times}$, and $\boldsymbol{g}_2=\boldsymbol{g}_3=0$, or 
(ii) $\boldsymbol{g}_2$ and $\boldsymbol{g}_3$ are i.i.d.~uniform over $\mathbbmss{F}_q^{\times}$, and $\boldsymbol{g}_1=\boldsymbol{g}_4=0$, with probability $1/2$ for each case. 
\item If $j=2$, then $\boldsymbol{g}_1,\boldsymbol{g}_2,\boldsymbol{g}_3$ are i.i.d.~uniform over $ \mathbbmss{F}_q^{\times}$, and $\boldsymbol{g}_4$ is conditionally uniformly distributed over ${ \mathbbmss{F}_q^{\times}\setminus \{g_2g_3/g_1\}}$ given $(\boldsymbol{g}_1,\boldsymbol{g}_2,\boldsymbol{g}_3)=(g_1,g_2,g_3)$. 
\end{itemize}

For this example, the user constructs a random $i=1$-sparse vector of length $2$ and a random $j=1$-regular invertible $2\times 2$ matrix, as defined earlier.

The user then constructs the coefficient vector $\mathbf{v}_1$ corresponding to the linear combination $\mathbf{Y}_1$,  
\[\mathbf{Y}_1 = [\boldsymbol{h}_1,\boldsymbol{h}_2]\cdot [\mathrm{X}_{c,1},\mathrm{X}_{d,1}]^{\mathsf{T}} = \boldsymbol{h}_1\mathrm{X}_{c,1}+\boldsymbol{h}_2\mathrm{X}_{d,1},\] 
which reduces to either $\boldsymbol{h}_1\mathrm{X}_{c,1}$ or $\boldsymbol{h}_2\mathrm{X}_{d,1}$ depending on which of $\boldsymbol{h}_1$ or $\boldsymbol{h}_2$ is set to be zero. 

Next, the user constructs the coefficient vectors $\mathbf{v}_2,\dots,\mathbf{v}_5$ corresponding to the linear combinations $\mathbf{Y}_2,\dots,\mathbf{Y}_5$, 
\begin{align*}
\mathbf{Y}_2 &= \mathbf{Y}_1+[\boldsymbol{g}_{1},\boldsymbol{g}_{2}]\cdot [\mathrm{X}_{a,1},\mathrm{X}_{b,1}]^{\mathsf{T}},\\
\mathbf{Y}_3 &= \mathbf{Y}_1+[\boldsymbol{g}_{3},\boldsymbol{g}_{4}]\cdot [\mathrm{X}_{a,1},\mathrm{X}_{b,1}]^{\mathsf{T}},\\
\mathbf{Y}_4 &= \mathbf{Y}_1+[\boldsymbol{g}_{1},\boldsymbol{g}_{2}]\cdot [\mathrm{X}_{a,2},\mathrm{X}_{b,2}]^{\mathsf{T}},\\
\mathbf{Y}_5 &= \mathbf{Y}_1+[\boldsymbol{g}_{3},\boldsymbol{g}_{4}]\cdot [\mathrm{X}_{a,2},\mathrm{X}_{b,2}]^{\mathsf{T}}.
\end{align*} 
 
For example, if $\boldsymbol{h}_2=0$ and $\boldsymbol{g}_{2}=\boldsymbol{g}_{3}=0$, then  
\begin{align*}
\mathbf{Y}_1 & = \boldsymbol{h}_1\mathrm{X}_{c,1},\\
\mathbf{Y}_2 & = \boldsymbol{g}_{1}\mathrm{X}_{a,1}+\boldsymbol{h}_1\mathrm{X}_{c,1},\\
\mathbf{Y}_3 & = \boldsymbol{g}_{4}\mathrm{X}_{b,1}+\boldsymbol{h}_1\mathrm{X}_{c,1},\\
\mathbf{Y}_4 & = \boldsymbol{g}_{1}\mathrm{X}_{a,2}+\boldsymbol{h}_1\mathrm{X}_{c,1},\\
\mathbf{Y}_5 & = \boldsymbol{g}_{4}\mathrm{X}_{b,2}+\boldsymbol{h}_1\mathrm{X}_{c,1},
\end{align*} which result in
\begin{align*}
\mathbf{v}_1 & = [0,0,0,0,\boldsymbol{h}_1,0,0,0],\\
\mathbf{v}_2 & = [\boldsymbol{g}_{1},0,0,0,\boldsymbol{h}_1,0,0,0],\\
\mathbf{v}_3 & = [0,0,\boldsymbol{g}_{4},0,\boldsymbol{h}_1,0,0,0],\\
\mathbf{v}_4 & = [0,\boldsymbol{g}_{1},0,0,\boldsymbol{h}_1,0,0,0],\\
\mathbf{v}_5 & = [0,0,0,\boldsymbol{g}_{4},\boldsymbol{h}_1,0,0,0].
\end{align*}

The collections $\{\mathbf{v}_1,\dots,\mathbf{v}_5\}$ and $\{\mathbf{Y}_1,\dots,\mathbf{Y}_5\}$ define a specific type of query set and its corresponding answer set, respectively, and different realizations of $\boldsymbol{h}_1$, $\boldsymbol{g}_{1}$, and $\boldsymbol{g}_{4}$ yield different realizations of this query/answer set type. 
Moreover, the probability of generating a query/answer set of this type is $1/15$. 
This is because the pair $(i,j)=(1,1)$ is selected with probability $4/15$, and, given this selection, $\boldsymbol{h}_2$, $\boldsymbol{g}_{2}$, and $\boldsymbol{g}_{3}$ are set to zero with probability $(1/2)\times (1/2) = 1/4$. 

Table~\ref{tab:1} lists all possible types of answer sets $\mathbf{Y}:=\{\mathbf{Y}_1,\dots,\mathbf{Y}_5\}$, with each row representing a distinct type and its corresponding probability $P_{\mathbf{Y}}$. 
For clarity, only the combination coefficients that are nonzero random variables are shown for each type, while those set to zero are omitted. 

{\renewcommand{\arraystretch}{1.65}
\begin{table*}[tb]
    \centering
    \caption{All possible types of answer sets and their corresponding probabilities.
    }
    \scalebox{1.075}{
    \begin{tabular}{@{}|@{}c@{}|@{}c@{}|@{}c@{}|@{}c@{}|@{}c@{}|@{}c@{}|@{}c@{}|@{}}
    \hline
    $\hspace{3pt}(i,j)\hspace{3pt}$ &
    $\mathbf{Y}_{1}$ & $\mathbf{Y}_{2}$ & $\mathbf{Y}_{3}$  & $\mathbf{Y}_{4}$ & $\mathbf{Y}_{5}$ &   $\hspace{3pt}P_{\mathbf{Y}}\hspace{3pt}$ \\  
    \hline
    $(0,1)$ &   $\emptyset$ &   $\boldsymbol{g}_{1}\mathrm{X}_{a,1}$ & $\boldsymbol{g}_{4}\mathrm{X}_{b,1}$ & $\boldsymbol{g}_{1}\mathrm{X}_{a,2}$  & $\boldsymbol{g}_{4}\mathrm{X}_{b,2}$  & $\frac{1}{15}$ \\
    \hline
    $(0,1)$ &   $\emptyset$ &   $\boldsymbol{g}_{2}\mathrm{X}_{b,1}$ & $\boldsymbol{g}_{3}\mathrm{X}_{a,1}$ & $\boldsymbol{g}_{2}\mathrm{X}_{b,2}$  & $\boldsymbol{g}_{3}\mathrm{X}_{a,2}$  & $\frac{1}{15}$ \\
    \hline
    $(0,2)$ &   $\emptyset$ &   $\boldsymbol{g}_{1}\mathrm{X}_{a,1}+\boldsymbol{g}_{2}\mathrm{X}_{b,1}$ & $\boldsymbol{g}_{3}\mathrm{X}_{a,1}+\boldsymbol{g}_{4}\mathrm{X}_{b,1}$ & $\boldsymbol{g}_{1}\mathrm{X}_{a,2}+\boldsymbol{g}_{2}\mathrm{X}_{b,2}$ & $\boldsymbol{g}_{3}\mathrm{X}_{a,2}+\boldsymbol{g}_{4}\mathrm{X}_{b,2}$  & $\frac{1}{15}$ \\
    \hline
    $(1,1)$ &  $\boldsymbol{h}_{1}\mathrm{X}_{c,1}$ &   $\boldsymbol{g}_{1}\mathrm{X}_{a,1}+\boldsymbol{h}_{1}\mathrm{X}_{c,1}$ & $\boldsymbol{g}_{4}\mathrm{X}_{b,1}+\boldsymbol{h}_{1}\mathrm{X}_{c,1}$ & $\boldsymbol{g}_{1}\mathrm{X}_{a,2}+\boldsymbol{h}_{1}\mathrm{X}_{c,1}$  & $\boldsymbol{g}_{4}\mathrm{X}_{b,2}+\boldsymbol{h}_{1}\mathrm{X}_{c,1}$  & $\frac{1}{15}$ \\
    \hline
    $(1,1)$ &  $\boldsymbol{h}_{1}\mathrm{X}_{c,1}$ &   $\boldsymbol{g}_{2}\mathrm{X}_{b,1}+\boldsymbol{h}_{1}\mathrm{X}_{c,1}$ & $\boldsymbol{g}_{3}\mathrm{X}_{a,1}+\boldsymbol{h}_{1}\mathrm{X}_{c,1}$ & $\boldsymbol{g}_{2}\mathrm{X}_{b,2}+\boldsymbol{h}_{1}\mathrm{X}_{c,1}$  & $\boldsymbol{g}_{3}\mathrm{X}_{a,2}+\boldsymbol{h}_{1}\mathrm{X}_{c,1}$  & $\frac{1}{15}$ \\
    \hline
     $(1,1)$ &  $\boldsymbol{h}_{2}\mathrm{X}_{d,1}$ &   $\boldsymbol{g}_{1}\mathrm{X}_{a,1}+\boldsymbol{h}_{2}\mathrm{X}_{d,1}$ & $\boldsymbol{g}_{4}\mathrm{X}_{b,1}+\boldsymbol{h}_{2}\mathrm{X}_{d,1}$ & $\boldsymbol{g}_{1}\mathrm{X}_{a,2}+\boldsymbol{h}_{2}\mathrm{X}_{d,1}$  & $\boldsymbol{g}_{4}\mathrm{X}_{b,2}+\boldsymbol{h}_{2}\mathrm{X}_{d,1}$  & $\frac{1}{15}$ \\
    \hline
    $(1,1)$ &  $\boldsymbol{h}_{2}\mathrm{X}_{d,1}$ &   $\boldsymbol{g}_{2}\mathrm{X}_{b,1}+\boldsymbol{h}_{2}\mathrm{X}_{d,1}$ & $\boldsymbol{g}_{3}\mathrm{X}_{a,1}+\boldsymbol{h}_{2}\mathrm{X}_{d,1}$ & $\boldsymbol{g}_{2}\mathrm{X}_{b,2}+\boldsymbol{h}_{2}\mathrm{X}_{d,1}$  & $\boldsymbol{g}_{3}\mathrm{X}_{a,2}+\boldsymbol{h}_{2}\mathrm{X}_{d,1}$  & $\frac{1}{15}$ \\
    \hline
    $(1,2)$ &  $\boldsymbol{h}_{1}\mathrm{X}_{c,1}$ &   $\boldsymbol{g}_{1}\mathrm{X}_{a,1}+\boldsymbol{g}_{2}\mathrm{X}_{b,1}+\boldsymbol{h}_{1}\mathrm{X}_{c,1}$ & $\boldsymbol{g}_{3}\mathrm{X}_{a,1}+\boldsymbol{g}_{4}\mathrm{X}_{b,1}+\boldsymbol{h}_{1}\mathrm{X}_{c,1}$ & $\boldsymbol{g}_{1}\mathrm{X}_{a,2}+\boldsymbol{g}_{2}\mathrm{X}_{b,2}+\boldsymbol{h}_{1}\mathrm{X}_{c,1}$ & $\boldsymbol{g}_{3}\mathrm{X}_{a,2}+\boldsymbol{g}_{4}\mathrm{X}_{b,2}+\boldsymbol{h}_{1}\mathrm{X}_{c,1}$  & $\frac{2}{15}$ \\
    \hline
    $(1,2)$ &  $\boldsymbol{h}_{2}\mathrm{X}_{d,1}$ &   $\boldsymbol{g}_{1}\mathrm{X}_{a,1}+\boldsymbol{g}_{2}\mathrm{X}_{b,1}+\boldsymbol{h}_{2}\mathrm{X}_{d,1}$ & $\boldsymbol{g}_{3}\mathrm{X}_{a,1}+\boldsymbol{g}_{4}\mathrm{X}_{b,1}+\boldsymbol{h}_{2}\mathrm{X}_{d,1}$ & $\boldsymbol{g}_{1}\mathrm{X}_{a,2}+\boldsymbol{g}_{2}\mathrm{X}_{b,2}+\boldsymbol{h}_{2}\mathrm{X}_{d,1}$ & $\boldsymbol{g}_{3}\mathrm{X}_{a,2}+\boldsymbol{g}_{4}\mathrm{X}_{b,2}+\boldsymbol{h}_{2}\mathrm{X}_{d,1}$  & $\frac{2}{15}$ \\
    \hline
    $(2,1)$ & $\boldsymbol{h}_1\mathrm{X}_{c,1}+\boldsymbol{h}_2\mathrm{X}_{d,1}$ &   $\boldsymbol{g}_1\mathrm{X}_{a,1}+\boldsymbol{h}_1\mathrm{X}_{c,1}+\boldsymbol{h}_2\mathrm{X}_{d,1}$ &   $\boldsymbol{g}_4\mathrm{X}_{b,1}+\boldsymbol{h}_1\mathrm{X}_{c,1}+\boldsymbol{h}_2\mathrm{X}_{d,1}$ &  $\boldsymbol{g}_1\mathrm{X}_{a,2}+\boldsymbol{h}_1\mathrm{X}_{c,1}+\boldsymbol{h}_2\mathrm{X}_{d,1}$ &  $\boldsymbol{g}_4\mathrm{X}_{b,2}+\boldsymbol{h}_1\mathrm{X}_{c,1}+\boldsymbol{h}_2\mathrm{X}_{d,1}$ &  $\frac{2}{15}$ \\
    \hline
    $(2,1)$ & $\hspace{4pt}\boldsymbol{h}_1\mathrm{X}_{c,1}+\boldsymbol{h}_2\mathrm{X}_{d,1}\hspace{4pt}$ &   $\hspace{4pt}\boldsymbol{g}_2\mathrm{X}_{b,1}+\boldsymbol{h}_1\mathrm{X}_{c,1}+\boldsymbol{h}_2\mathrm{X}_{d,1}\hspace{4pt}$ &   $\hspace{4pt}\boldsymbol{g}_3\mathrm{X}_{a,1}+\boldsymbol{h}_1\mathrm{X}_{c,1}+\boldsymbol{h}_2\mathrm{X}_{d,1}\hspace{4pt}$ &  $\hspace{4pt}\boldsymbol{g}_2\mathrm{X}_{b,2}+\boldsymbol{h}_1\mathrm{X}_{c,1}+\boldsymbol{h}_2\mathrm{X}_{d,1}\hspace{4pt}$ &  $\hspace{4pt}\boldsymbol{g}_3\mathrm{X}_{a,2}+\boldsymbol{h}_1\mathrm{X}_{c,1}+\boldsymbol{h}_2\mathrm{X}_{d,1}\hspace{4pt}$ & $\frac{2}{15}$  \\
    \hline
    \end{tabular}}
    \label{tab:1}
\end{table*}
}

The user randomly generates a realization $\{\mathrm{v}_1,\dots,\mathrm{v}_5\}$ of $\{\mathbf{v}_1,\dots,\mathbf{v}_5\}$ and sends the vectors 
$\mathrm{v}_1,\dots,\mathrm{v}_5$ to  servers $1,\dots,5$ in a randomly chosen order. 
The servers then return the corresponding linear combinations $\mathrm{Y}_1,\dots,\mathrm{Y}_5$ to the user.   

Upon receiving $\mathrm{Y}_1,\dots,\mathrm{Y}_5$, the user then recovers $\mathrm{X}_{a,1},\mathrm{X}_{a,2}$ and $\mathrm{X}_{b,1},\mathrm{X}_{b,2}$. 
For instance, in our example, the user recovers $\mathrm{X}_{a,1}$ as ${(\mathrm{Y}_2-\mathrm{Y}_1)/g_1}$, $\mathrm{X}_{a,2}$ as ${(\mathrm{Y}_4-\mathrm{Y}_1)/g_1}$, $\mathrm{X}_{b,1}$ as ${(\mathrm{Y}_3-\mathrm{Y}_1)/g_4}$, and $\mathrm{X}_{b,2}$ as ${(\mathrm{Y}_5-\mathrm{Y}_1)/g_4}$.   

Similarly, it can be verified that the user can recover $\mathrm{X}_{a,1},\mathrm{X}_{a,2}$ and $\mathrm{X}_{b,1},\mathrm{X}_{b,2}$ from any realization of any type of answer set presented in Table~\ref{tab:1}. 
This confirms that the recoverability condition is satisfied. 

To show that the privacy condition is also satisfied, we prove the following: 
\begin{itemize}
\item[(i)] If a server is requested to send a single subpacket to the user, the probability that it belongs to a demand message is equal to the probability that it belongs to an interference message.
\item[(ii)] If a combination of two subpackets is requested, the probabilities of these subpackets belonging to two demand messages, one demand and one interference, and two interference messages are all equal.
\item[(iii)] If a combination of three subpackets is requested, the probability that they belong to one demand and two interference messages is equal to the probability that they belong to one interference and two demand messages.
\end{itemize}

\vspace{0.125cm}
\subsubsection*{Proof of (i)} 
From Table~\ref{tab:1}, if a single subpacket is requested, it belongs to $\mathrm{X}_a$ (or $\mathrm{X}_b$) with probability $2\times 2 \times (1/5) \times (1/15) = 4/75$, as the subpacket can be either $\mathrm{X}_{a,1}$ or $\mathrm{X}_{a,2}$ (or $\mathrm{X}_{b,1}$ or $\mathrm{X}_{b,2}$), and each of these subpackets appears in two rows, with each of these rows having a probability of $1/15$.

Similarly, the probability that the requested subpacket belongs to $\mathrm{X}_c$ (or $\mathrm{X}_d$) is $2 \times (1/5) \times (1/15) + (1/5) \times (2/15) = 4/75$, as $\mathrm{X}_{c,1}$ (or $\mathrm{X}_{d,1}$) appears in three rows, two of which have a probability of $1/15$ and the other one has a probability of $2/15$.

\vspace{0.125cm}
\subsubsection*{Proof of (ii)} 
Again, from Table~\ref{tab:1}, if a combination of two subpackets is requested, these subpackets belong to $\mathrm{X}_a$ and $\mathrm{X}_b$ with probability $2 \times (2/5) \times (1/15) = 4/75$. 
This is because the two subpackets can either be $\mathrm{X}_{a,1}$ and $\mathrm{X}_{b,1}$ or $\mathrm{X}_{a,2}$ and $\mathrm{X}_{b,2}$, with each combination appearing twice in a single row, which has a probability of $1/15$. 

Similarly, the probability that one of the two subpackets in the requested combination belongs to $\mathrm{X}_a$ (or $\mathrm{X}_b$) and the other one belongs to $\mathrm{X}_c$ (or $\mathrm{X}_d$) is $2\times 2\times (1/5) \times (1/15) = 4/75$. 
This is because one of the two subpackets can be either $\mathrm{X}_{a,1}$ or $\mathrm{X}_{a,2}$ (or $\mathrm{X}_{b,1}$ or $\mathrm{X}_{b,2}$) and the other one can be either $\mathrm{X}_{c,1}$ or $\mathrm{X}_{c,2}$ (or $\mathrm{X}_{d,1}$ or $\mathrm{X}_{d,2}$), with each  combination appearing in two different rows, each of which has a probability of $1/15$. 

Finally, the two subpackets in the requested combination belong to $\mathrm{X}_c$ and $\mathrm{X}_d$ with probability $2 \times (1/5) \times (2/15) = 4/75$. 
This is because a combination of
$\mathrm{X}_{c,1}$ and $\mathrm{X}_{d,1}$ appears in two rows, each of which has a probability of $2/15$.

\vspace{0.125cm}
\subsubsection*{Proof of (iii)}
From Table~\ref{tab:1}, if a combination of three subpackets is requested, the probability that one of these subpackets belongs to $\mathrm{X}_a$ (or $\mathrm{X}_b$) and the other two subpackets belong to $\mathrm{X}_c$ and $\mathrm{X}_d$ is $2 \times 2\times (1/5) \times (2/15) = 8/75$.
This is because one of the three subpackets can be either $\mathrm{X}_{a,1}$ or $\mathrm{X}_{a,2}$ (or $\mathrm{X}_{b,1}$ or $\mathrm{X}_{b,2}$) and the other two must be $\mathrm{X}_{c,1}$ and $\mathrm{X}_{d,1}$, with each  combination appearing in two rows, each of which has a probability of $2/15$.

Similarly, one of the three subpackets in the requested combination belongs to $\mathrm{X}_c$ (or $\mathrm{X}_d$) and the other two subpackets belong to $\mathrm{X}_a$ and $\mathrm{X}_b$ with probability $2\times (2/5) \times (2/15) = 8/75$.
This is because one of the three subpackets can be $\mathrm{X}_{c,1}$ (or $\mathrm{X}_{d,1}$) and the other two can either be $\mathrm{X}_{a,1}$ and $\mathrm{X}_{b,1}$ or $\mathrm{X}_{a,2}$ and $\mathrm{X}_{b,2}$, with each  combination appearing twice in a single row, which has a probability of $2/15$.  

\vspace{0.125cm}
For this example, the expected number of bits downloaded from all servers (normalized by the number of bits per message) is equal to 
$2\times (3/15) + (5/2)\times (12/15) = 12/5$, 
which gives a rate of $R = 2/(12/5)=5/6$.
This matches the capacity $C$ in~\eqref{eq:C} for $K=4$, $D=2$, and $N=5$. 

\bibliographystyle{IEEEtran}
\bibliography{PIR_PC_Refs}

\end{document}